# MAXIMIZING NETWORK CAPACITY, CONTROL AND MANAGEMENT IN DESIGNING A TELEMEDICINE NETWORK: A REVIEW AND RECENT CHALLENGES

**AUTHORS:**
B. O. Sadiq[1], O. S. Zakariyya[2], M. D. Buhari[3], and A. N. Shuaibu[4]

**AFFILIATIONS:**
[1,3,4]Department of Electrical, Telecommunication and Computer Engineering, Kampala International University, Uganda.
[1]Department of Computer Engineering, Ahmadu Bello University, Zaria, Nigeria.
[2]Department of Electrical and Electronic Engineering, University of Ilorin, Kwara State Nigeria.
[3]Department of Electrical and Electronic Engineering, ATBU Bauchi, Bauchi State Nigeria.
[4]Department of Electrical and Electronics Engineering, University of Jos, Jos Nigeria.

**\*CORRESPONDING AUTHOR:**
Email: bosadiq@kiu.ac.ug

**ARTICLE HISTORY:**
**Received:** 21 December, 2023.
**Revised:** 12 February, 2024.
**Accepted:** 07 March, 2024.
**Published:** 31 March, 2024.

**ARTICLE INCLUDES:**
Peer review

**DATA AVAILABILITY:**
On request from author(s)

**EDITORS:**
Ozoemena Anthony Ani

**FUNDING:**
None

## Abstract

*Telemedicine networks have seen significant changes in their capacity, monitoring, management, and control framework during the previous decades. The evolution of network capacity, control, and management for Unmanned Aerial Vehicle (UAV) & Software-Defined Networks (SDN) as support to telemedicine, artificial intelligence in telemedicine networks, and capabilities in designing a telemedicine network with respect to its performance and customization is presented in this study, with a historical history and a future view. The first section of the article goes over the history of traffic and capacity expansion, as well as future projections. By introducing a medical and image data communication protocol for telemedicine, the second section examines the technological constraints of expanding capacity in the era of UAV & software-defined networking. The third section discusses ways to maximize network capacity by considering quality of service (QoS) capacity issues. Finally, the article explores how to construct a telemedicine network that can provide performance, customization, and capabilities to keep up with increased traffic in the coming decades. Research gaps and future directions were presented in the last section.*

## 1.0 INTRODUCTION

Advancements in the telemedicine industry have been with the use of information and communication technology (ICT) and its tools. These ICT tools are used not only for visual communication but also for patient data transfer and collaborative learning. Nonetheless, achieving these uses needs a communication medium that can provide high spectral efficiency. As such, this review paper draws contribution from works that delves into the historical evolution of traffic and capacity, both retrospectively and prospectively, papers that used UAV and SDN to improve telemedicine network by considering medical and image data communication protocol and WBAN framework, and those that focus on enhancing telemedicine network using QoS mechanism, machine learning and cloud infrastructure. This will therefore improve telemedicine network design capable of delivering performance, adaptability, and capacities essential for accommodating the anticipated surge in traffic in the upcoming decades.

The need for high spectral efficiency to increase fiber utilization and the availability of channels with different bandwidth requirements have been the main drivers for Optical Networks (ONs). However, these drivers have been influenced by the nature of their management, capacity, and control. Telemedicine currently uses optical networks which are high-capacity networking in the era of searching novel dimensions to increase transport capacity [1,2]. This is due to the fact that optical networks have seen significant changes in their capacity, monitoring, management, and control framework during the previous two decades. Due to recent advancements in optical solutions that lead to real-time patient monitoring, better emergency care, secure data transmission, improved patient access, and collaboration between the medical team and members, the telemedicine industry has adopted the use of this network with advanced body sensors technologies such as an electrocardiogram (EKG), electroence-phalography (EEG), heart rate, body temperature and monitoring and assistance to patients with chronic diseases. These are however achieved using wired and wireless infrastructures, real-time data processing, and interactive interfaces [3,4].

According to the medical association that uses telehealth, rural health points have been able to connect using optical networks with the help of connected medical equipment ranging from simple scales, blood pressure monitors, and thermometers for primary parameters to medical imaging equipment such as microscopes, scanners, magnetic resonance imaging, (MRI) and X-rays [5-8]. The authors of [9] have discussed the prospects for the growth of e-health through the integration of optical networks.

Telemedicine was originally introduced to the world in the early 1960s. It first emerged in metropolitan areas before making its way into the arena of emergency medicine. Telemedicine grew in popularity in rural areas, where people with limited or no access could now consult professionals from a distance. In 1967, the University of Miami School of Medicine was one of the first schools to collaborate with the local fire department to send electrocardiographic rhythms to Jackson Memorial Hospital through radio. Large-scale wired and wireless networks, as well as mobile computing technologies such as optical networks, Wi-Fi mesh, WiMAX, and cellular 3G among others, allow healthcare personnel to access important information from inside healthcare network systems at any time and from any place. Presently, the coming of telemedicine with advancement in network research had made it more prominent in carrying out telemedicine key applications like store and forward, remote patient monitoring, and interactive telemedicine/telehealth [10-12]. Patients and physicians can communicate directly via interactive telemedicine/telehealth. These sessions can take place at the patient's home or a medical kiosk. Video conferencing applications may be used in the interactions. The store and forward method, also known as asynchronous telemedicine, allows the healthcare provider to transmit patient data with another, such as lab results amongst others [13]. Remote patient monitoring, on the other hand, allows patients to be observed at home using handheld apps or remote sensors that collect data on temperature, blood sugar levels, blood pressure, and other vital indicators. Telemonitoring is the term for this [14].

Optical networks have been used in telemedicine as one of the most promising types of networks [15]. Previous research, on the other hand, relied on a wireless mesh network. Wireless mesh networks are utilized to provide a stable and reliable communications infrastructure backbone for telemedicine systems. They are chosen over other wireless networks because the broadband services they provide are stable. in addition, they are also robust because if one or more mesh nodes fail, the other mesh nodes may still route the data. Wireless mesh networks not only extend coverage as well as save costs on wiring and personnel management. Interference and fluctuating load, however, diminish their performance [16]. In wireless mesh networks, interference is a major problem. Because of the overcrowded unlicensed radio frequency spectrum, interference arises. A solution to this is the use of Optical Mesh Networks (OMN). In mesh network design, an optical mesh network is a type of optical telecommunications network that uses either wired fiber-optic transmission or wireless free-space optical communication. In wired optical networks, light signals go through a fiber, but in free-space optical communication, light signals travel over the air [17,18].

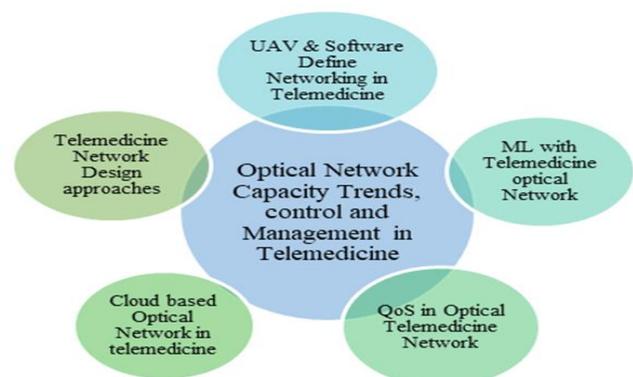

**Figure 1:** Different aspects of Optical Network Capacity Trends and Management in Telemedicine

A graphical view of the various aspects of research that contributes to the increase capacity and management of optical networks when applied to telemedicine applications is presented in Figure 1.

Significant research has been done on how to increase the capacity, control, and management of optical networks to serve bandwidth-dependent applications like telemedicine. The earlier approaches involved the continuous manufacturing of optical fibers that can carry more traffic [19]. The process of the manufactured fiber was subsequently improved using different usage optimization techniques in several ways like routing protocol design [20], Quality of Service (QoS) [21], software-defined networking [22], machine learning [23], etc. These optimization techniques served as tools to improve capacity in terms of reducing delay, preventing packet loss amongst others, whilst managing and controlling the optical networks with a view to increasing their capacities. Another area of research is cloud-based optical networks. This type of network combines the power of optical network technologies and cloud computing. This type of network is utilized to provide telemedicine applications with high-speed local networks [24,25]. For high-speed, high-quality, and high-capacity internet traffic, [26] proposes an impairment-aware reactive defragmentation strategy for dynamic elastic optical networks servicing telehealth care-oriented traffic. The suggested technique was designed to provide high-priority data on the shortest path available between the source and destination nodes, minimizing latency.

There are several research efforts on Internet of Things (IoT) applications with telemedicine [27,28], with a particular emphasis on high-speed connectivity, with the progress of sensor technology and connectivity. The application of artificial intelligence (AI) and machine learning (ML) techniques in this field allows for the gathering, analysis, and real-time decision-making process of patient data [29,30]. For example, in the study [31], the authors looked at the movement of heart rate in the context of internet of things, telemedicine and mobile health. Optimization of a remote public medical emergency management system based on the internet of things [32,33] has also been studied in a number of studies recently with little latency. However, the aforementioned applications are fraught with difficulties. Dealing with a variety of protocols necessitates specialist tools (scalability), and internet disruption can kill a patient, implying that network failure is not acceptable, necessitating a high-speed capacity network. Other strategic concerns include, for example, government policies on patient record security and environmental hazards.

Optical network capacity and trends are highlighted in many review articles as such as [31] as a promising subject for future research. The highlighted subject follows the typical telemedicine center framework presented in Figure 2.

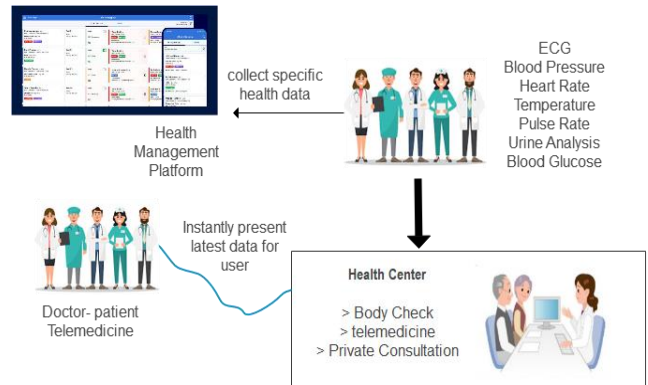

**2(a):** Internal Structure

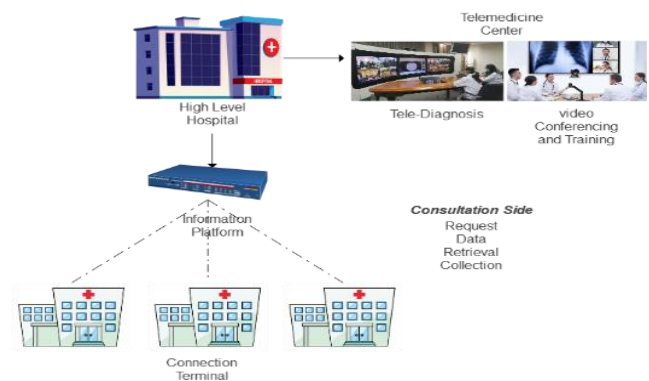

**2(b):** Connecting Infrastructure
**Figure 2:** A typical Telemedicine Center

A well-written review article describes the past and current state of the topic, offers future directions, and summarizes the previous development of the issue. Several studies on the optical network, control, and management have been written in various domains, according to the literature. Researchers give review articles on the use of optical networks in UAV networks, agricultural modernization, disaster management, remote sensing, communication networks, and power transmission line monitoring, among other topics.

The following are the main features of this paper:
- It is the first review study based on capacity trends, control, and management of the usage of optical networks in designing a telemedicine network.

- It examines several domains of telemedicine where optical networks are employed, as well as a comprehensive literature review on the design issue.
- It emphasizes the larger features as well as the parallel research directions employed to improve telemedicine network capacity.
- It outlines the research shortcomings in the discipline and the areas where more research is required.
- It identifies the many types of issues that the telemedicine field faces and lists the remedies suggested by various researchers.

Therefore, the evolution of optical network capacity and management to support telemedicine networks, software-defined networks as support to telemedicine, machine learning in optical networks, UAV-based telemedicine network, and capabilities in designing a telemedicine network with respect to its performance and customization is presented in this study, with a historical history and a future view.

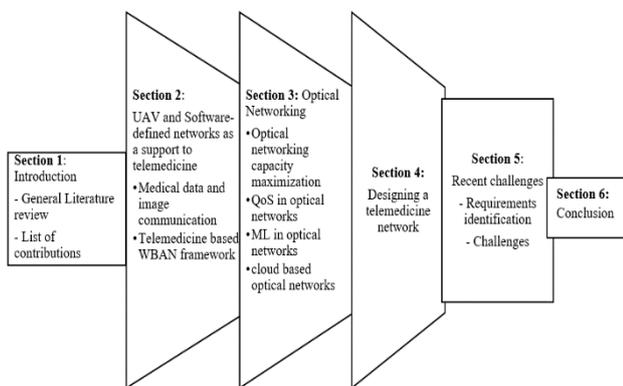

**Figure 3:** The Detailed structure of the review process.

The paper's organization is as follows: it commences with a broad introduction to the subject, encompassing recent advancements in optical networks within the telemedicine sector, the diverse range of applications in telemedicine networks, and the imperative for employing high-capacity optical networks. The subsequent subsections delve deeper. The first subsection delves into the historical evolution of traffic and capacity, both retrospectively and prospectively. In the second section, a medical and image data communication protocol for telemedicine is introduced, scrutinizing the technological limitations of capacity expansion in the context of software-defined networking. Moving forward, the third section explores strategies for optimizing network capacity, considering the challenges of maintaining quality of service (QoS). Ultimately, the paper investigates the construction of a telemedicine network capable of delivering performance, adaptability, and capacities essential for accommodating the anticipated surge in traffic in the upcoming decades. A visual representation detailing the comprehensive structure of the review process is illustrated in Figure 3.

## 2.0 UAV AND SOFTWARE-DEFINED NETWORKS AS SUPPORT TO TELEMEDICINE

By offering a dependable and secure way to transfer medical data and supplies to remote or disaster-affected locations, Unmanned Aerial Vehicles (UAVs) [34,35] and Software-defined Networks (SDNs) [36-37] can enhance telemedicine in a significant way. UAVs can be used to deliver medical supplies and equipment to far-flung locations that are challenging to access by conventional transportation. Also, they can be utilized to deliver medical samples from these regions back to labs for examination and analysis. Additionally, by enabling doctors to remotely monitor vital signs and converse with patients via video conferencing technology, UAVs can be utilized to deliver real-time medical help to patients in remote locations. On the other hand, SDNs can promote telemedicine by giving healthcare practitioners and patients a safe and dependable way to transfer medical information. SDNs enable the development of virtual networks that may be dynamically rearranged to satisfy the varying requirements of telemedicine applications [37]. So, even in remote or disaster-affected locations, healthcare professionals can give patients more individualized care.

UAVs and SDNs can work together to create a potent technology stack that will assist telemedicine and enhance patient outcomes in underserved areas. Yet, it's critical to make sure that these technologies are used responsibly and ethically, considering issues with effective communication, security, privacy, and other moral issues. The issues of effective communication, security, and privacy amongst others can be achieved via the development of an efficient communication protocol and framework. Therefore, some research works such as [38,39] have presented communication protocols that consider effective medical and image data communication. Therefore, the success of increasing the capacity, control, and management of optical networks in telemedicine networks can also be dependent on the communication protocol and connectivity framework. A conceptual diagram of how UAVs can enhance the telemedicine network [40] is depicted in Figure 4. Nonetheless, the success of the framework is dependent on the communication protocol used.

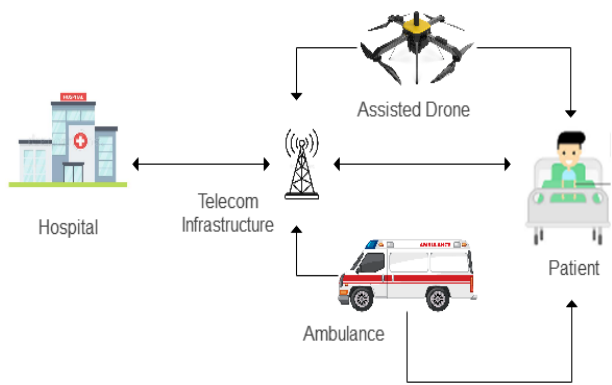

**Figure 4:** UAV-based telemedicine network

## 2.1 Medical data and image communication protocol

With a view to maximizing the capacity of a telemedicine network, some authors such as the work [41,43] have developed medical data and image communication protocols as well as discussed, the communications requirements for image data in telemedicine. Based on the gathered perception, text, pictures, and sound must be sent and received for telemedical applications. The text contains patient data, laboratory results, and ECG (heart tracing) results. Still, photos and full-motion videos are both examples of images. Further to voice and chest sounds, radiological pictures, slides, and graphics may also be communicated. As presented by [41], up to two or three simultaneous video bitstreams could be needed for telemedicine systems: two low-rate bitstreams for teleconferencing and an optional high-rate bitstream for diagnostic video. To achieve communication, use H.261 (64 kbps to 1.92 Mbps), H.263 (15-34 kb/s), or MPEG-1 (1.2-2 Mb/s). Nevertheless, the jitter and latency of real-time video streams are always important and should be reduced. The authors in the work of [42] also showed the capability of delivering medical ultrasound images over a wide area network in real-time. [43] from their work have stressed that the communication of images in telemedicine networks required much bandwidth. Therefore, there should be an efficient and reliable control mechanism [44]. As an alternative, a network of satellite ground stations with symmetric bandwidth connected by satellite is the architecture that was suggested. Other suggestions by earlier authors of telemedicine with respect to the topic of the presentation are presented in Table 1.

**Table 1:** Comparison of the articles on medical data and image communication protocol to improve telemedicine network

| Topic | Reference | Year of publication |
|---|---|---|
| Digital Imaging and Communications in Medicine (DICOM) | [45] | 2010 |
| Health Level 7 (HL7) Protocol | [46] | 1993 |
| Picture Archiving and Communication Systems (PACS) | [47] | 1994 |
| Simple Object Access Protocol (SOAP): | [48] | 2010 |
| Network capacity, control, and management protocol in Telemedicine | No record | |

In the work of [45], the use of Digital Imaging and Communications in Medicine (DICOM) as a standard in medical imaging was presented. The benefits of employing DICOM, such as enhanced workflow, improved interoperability, and increased efficiency in the sharing of patient and medical picture data, were also discussed. The authors also explain a number of real-world uses for DICOM in medical imaging, such as radiology, cardiology, and ophthalmology. The study [46] outlines the benefits of HL7, such as enhanced workflow, improved interoperability, and increased effectiveness in the interchange of medical data. The work also discusses the different parts of the HL7 protocol, including messages, segments, fields, and tables. Also, the author talks about how to use HL7 in healthcare contexts for things like clinical documentation, reporting test findings, and billing and claims processing. An overview of PACS technology, including its design, parts, and uses, was given in the work by [47]. The author analyzes the many PACS options, including standalone PACS, mini-PACS, and integrated PACS, as well as their benefits and drawbacks. The author also discusses how PACS is used in radiology, cardiology, and other fields of medicine.

The study also covers the difficulties and restrictions of putting PACS into practice, such as the requirement for standardization of data formats and protocols, security and privacy issues, and the high expense of putting PACS into practice and maintaining it. An overview of the SOAP protocol and how it is used to secure data transmission over the internet can be found in [48]. The use of SOAP in healthcare information systems is examined by the authors, along with its benefits and drawbacks, including its capacity to offer a secure and dependable means of data transfer. Additionally, the paper discusses the security prerequisites for EMR databases, such as confidentiality, integrity, and availability, and looks at how SOAP might be utilized to satisfy these prerequisites. The research argues that the SOAP protocol is appropriate for safeguarding the transmission of EMR databases since it offers a dependable and secure means of data transfer. Nevertheless, they also point out that adopting SOAP has several drawbacks, including its complexity and the requirement for additional security precautions to guarantee data confidentiality and integrity.

Generally, the summary of what each of the protocols represents is discussed as follows: Healthcare providers frequently exchange clinical and administrative data using the messaging protocol HL7. It is commonly utilized in healthcare IT applications and electronic health record (EHR) systems [49]. HL7 can be used in a range of healthcare situations, including telemedicine, and facilitates the sharing of numerous sorts of data, including medical pictures. The complexity and lack of capability for real-time data sharing are just two of HL7's drawbacks. In hospitals and clinics, PACS is a medical imaging technology that is frequently employed. It enables digital medical picture storage and retrieval and can be utilized in telemedicine applications to streamline the exchange of medical images between healthcare professionals [50]. PACS is generally easier to use than DICOM (digital Imaging and Communications in Medicine) and is accessible through regular web browsers.

However, PACS can be expensive and may require extensive IT infrastructure to implement. On the other hand, a common protocol used in telemedicine called DICOM is used to exchange medical images and related data [51]. It can be utilized in telemedicine applications that include the transmission of medical pictures and is commonly employed in radiology and other medical imaging disciplines. DICOM supports a wide range of image types, including X-rays, CT scans, and MRIs, and allows for the exchange of image data in a standardized format. Yet to understand and show pictures via DICOM, specialist software is needed. Another type of common protocol used in the telemedicine field to transfer medical data from source to destination is SOAP [52]. To make the interchange of medical data, including clinical data and medical imaging, easier, it can be employed in telemedicine applications. Several programming languages support SOAP, which is rather easy to implement. However, SOAP might not be as effective as other protocols, and it might need more bandwidth to send big volumes of data.

Based on the review of the medical data and image communication protocol, it can be concluded that the telemedicine network can be categorized into three different network layers which are the transportation protocol layer, the medical device layer, and the application layer. Most of the previously used telemedicine protocols were adopted rather than created as an integral part of medical data. It was determined that two types of protocol exist which are:
a. The Store-and-forward type of protocol
b. The real-time protocol

The earlier works in medicine use the store and forward type of protocol to control and manage the telemedicine applications. However, with recent advancements in the use of UAV and software-defined networking as support for telemedicine, the adopted protocols are Real-time. These Realtime protocols are typically used under the coined name of Wireless Body Area Network (WBAN) framework for telemedicine.

## 3.0 TELEMEDICINE-BASED WBAN FRAME WORK

This subsection presented the telemedicine based WBAN framework design as a contribution to improving the network capacity, control, and management in a telemedicine network. By enabling real-time remote patient monitoring, Wireless Body Area Networks (WBANs) are a promising technology for enhancing telemedicine networks. A number of wireless sensors are attached to the patient's body as part of the WBAN architecture, and these sensors gather information about the patient's temperature, blood pressure, and heart rate [53]. Then, a central hub, which may be situated in the patient's home or a healthcare facility, receives this data wirelessly. The healthcare provider can then use the data to monitor the patient's health status and deliver the necessary care after the hub sends it to them in real-time [54]. WBANs are subject to a number of standards, including IEEE 802.15.6 and ISO/IEEE 11073 [55,56]. These guidelines aid in ensuring interoperability among various hardware and software systems by defining the communication protocols and data formats used in WBANs. Standards-compliant WBANs can assist improve the dependability and security of telemedicine networks. WBANs can be used in telemedicine for a variety of purposes, such as the remote monitoring of chronic conditions like diabetes and hypertension, the home monitoring of elderly patients, and the monitoring of patients in intensive care units. WBANs can provide real-time patient monitoring, allowing medical professionals to react to changes in the patient's health status immediately.

WBANs have a few issues that need to be fixed despite any potential benefits they may have. The limited range and bandwidth of wireless communications, the requirement for frequent sensor battery replacement, and the possibility of interference from other wireless devices are a few of these drawbacks. furthermore, there are problems with the safety and confidentiality of patient information transferred over a wireless connection [57]. Some pertinent works that have used the WBAN framework to improve the telemedicine network are presented in Table 2.

**Table 2:** Summary of the articles on improving telemedicine network using WBAN framework

| Topic | Reference | Year of publication |
|---|---|---|
| Wireless body area network for telemedicine | [58] | 2020 |
| E-health beyond telemedicine | [59] | 2019 |
| Health telemonitoring | [60] | 2019 |
| WBAN-based remote monitoring system | [61] | 2023 |
| Real-time framework for patient monitoring | [62] | 2020 |
| Telemedicine framework in theCOVID-19 pandemic | [63] | 2022 |
| Network capacity, control, and management protocol in Telemedicine | No record | |

The use of Wireless Body Area Network (WBAN) technology in telemedicine for emergency care is covered in the study [58]. WBANs are wireless networks of medical sensors that track vital signs and other health-related information when they are worn on or implanted into a person. The benefits and drawbacks of adopting WBANs for emergency treatment are discussed in the study, including its capacity to deliver real-time monitoring and diagnosis, lower hospitalization expenses, and enhance patient outcomes. The authors also address the technological and legal concerns that must be resolved before WBAN-based telemedicine systems can be put in place. The report concludes that although WBAN-based telemedicine offers significant potential to enhance emergency care and save healthcare costs, more study and development are required to address the technological and legal issues. In [59], Wireless Body Area Networks (WBANs) are introduced and their potential to advance e-healthcare beyond telemedicine to remote health monitoring is explored.

Also, the article provides a review of the literature on WBANs and healthcare applications, such as remote patient monitoring and the management of chronic diseases. The authors also discuss the technological difficulties and solutions related to the implementation of WBAN-based healthcare systems, including energy-efficient protocols and security procedures. The report ends with a discussion of the potential implications for e-healthcare of the future paths of WBAN research concerning telemedicine. In [60], a novel method for secure patient data transmission across Wireless Body Area Networks (WBANs) for remote monitoring of health is proposed. To ensure a high level of data security during transmission, the authors advise employing a hybrid encryption strategy that combines symmetric and asymmetric encryption algorithms. To guard against unwanted access to and tampering with patient data, the proposed system additionally incorporates a method for patient authentication and data integrity verification. The authors presented a technique of using a wireless sensor network to track patient health information in real-time, such as vital signs and activity levels. The suggested method was said to increase the effectiveness and precision of remote health monitoring while simultaneously guaranteeing the privacy and security of patient data.

The author of [61] proposed a Wireless Body Area Network (WBAN)-based remote monitoring system for healthcare services using machine learning techniques. The system is intended to gather data from numerous WBAN sensors worn by patients, analyze that data using machine learning algorithms, and then deliver insights into the health status of those patients. A centralized server for data processing and storage is part of the suggested system, along with a user interface for viewing and analyzing patient data for healthcare professionals. It explored the system's technical difficulties, such as data security, privacy, and data processing capabilities. They also emphasized the potential advantages of the suggested method, such as enhanced diagnostic precision, lower healthcare expenses, and better patient outcomes. The research by [62] proposes a wireless body area network-based real-time platform for patient monitoring systems (WBAN). The proposed framework attempts to address the requirement for ongoing patient monitoring in medical facilities and at home.

The WBAN, which consists of a set of wearable sensors placed on the patient's body to collect physiological data, and a remote monitoring system that receives and interprets the acquired data, make up the framework's two primary parts. The ZigBee and Bluetooth communication protocol were suggested to be used by the authors to wirelessly transfer data from sensors to the remote monitoring system. The authors used a machine learning algorithm to categorize and forecast patient health status to deal with the massive amounts of data supplied by the WBAN. The results demonstrate that the proposed framework is capable of accurately detecting and classifying various medical conditions in real-time using a set of physiological signals collected from real patients.

The authors of [63] suggested a telemedicine framework for use in the COVID-19 pandemic. Due to the risk of infection, the framework is intended to offer remote healthcare services to patients who are unable or unwilling to visit hospitals or clinics. The authors talk about the difficulties experienced by healthcare professionals during the pandemic, such as the requirement to preserve social distance and lower the danger of virus exposure. They contend that telemedicine can be a practical means of delivering healthcare remotely while lessening the strain on medical institutions. A web-based platform for teleconsultations, remote monitoring tools, and electronic medical records are some of the components of the proposed system. The authors also examine the technological and legal difficulties that come with putting the framework into practice Eventually, they draw the following conclusions: Telemedicine can be crucial in the control of COVID-19 and other pandemics, and the suggested framework can serve as a useful guide for politicians and healthcare professionals.

Based on these works, it is evident that the WBAN-based framework for telemedicine has contributed to the efficient delivery of medical care over network infrastructure. However, to achieve seamless communication between patients, medical practitioners and other assisted telemedicine support such as the UAV and software-defined networks amongst others, there is a need to use a communication medium that guarantees the quality of service (QoS) and bandwidth utilization.

## 4.0    OPTICAL NETWORKING
Before now, certain telemedicine applications have been implemented using both basic telephone lines and wired communications technologies like ISDN and ADSL. However, mobile and ambulatory telemedicine systems can now be used alongside contemporary technologies like WLAN, Bluetooth, UMTS, GPRS, and Edge as well as satellite connectivity. The communication networks used with the telecommunication systems are presented in Table 3.

**Table 3:** Communications networks for telemedicine

| Network Type | Connection | Data Transfer and Frequency Band | Telemedicine Services |
|---|---|---|---|
| Satellite | Skybridge, Celestri ICO Global Star, amongst others | Ku band (16Kb/s-2MB/s), Ka band (155MB/s) amongst others. The different satellite has different bands and data transfer rates | voice, data and video transmission |
| GSM and UMTS (3G) | GSM-1900 (1900 MHz)/ UMTS | 9.6-43.4KB/s / 2Mbps | GSM supports Emergency telemedicine patient monitoring in Adhoc locations while UMTS can support high bandwidth mobile telemedicine applications |
| Wireless LAN | Bluetooth | 1Mbps and 2.4GHz | Home hospital patient tele monitoring and access to medical data in the hospital environment |
| | Home RF 1.0 | 1Mbps and 2.4GHz | |
| | Home RF 2.0 | 10Mbps and 2.4GHz | |
| | IEEE 802.11a/b/g/n | IEEE 802.11a: 54Mbps and 5GHz IEEE 802.11b: 11Mbps and 2.4GHz IEEE 802.11g: 54Mbps and 2.4GHz IEEE 802.11n: 600Mbps and 2.4GHz & 5GHz | |
| LTE(4G) and 5G | Broadband | 150Mbps and 20Gbps | High bandwidth applications and scenarios |
| 6G | Broadband | | High bandwidth applications and scenario |

It is important to note that various telemedicine applications have different communications technology needs in terms of complexity and range. There are no theoretical bandwidth requirements for the transmission of medical data. Longer transmission times are assumed when there is insufficient bandwidth. As such, there is a need to minimize longer transmission times. The telecommunication infrastructure in recent times that supports telemedicine applications makes use of an optical network as a backbone [64,66] in addition to linking different hospital centers in different locations [67]. This can be attributed to the high availability of bandwidth they possess. As such, there is a need to maximize its capacity.

Conclusively, we can argue that optical networks are crucial to telemedicine because they provide fast, dependable, and secure communication between healthcare professionals and patients.

### 4.1    Optical Network Capacity Maximization
To fulfill the rising demand for high-bandwidth applications and services, network operators must be able to maximize optical network capacity, which is a crucial component of network design. Some of the general techniques that can or have been used to maximize the network capacity of optical networks in telemedicine are discussed as follows:

a. Quality of Service (QoS)

To increase network performance, reliability, and user satisfaction in optical networks, quality of service (QoS) is a crucial approach. With the help of QoS, the network can give some traffic types priority over others, ensuring that vital applications and services have access to the bandwidth, latency, and dependability they need to operate at their best. Traffic Classification and Prioritization, Bandwidth Reservation, Traffic Shaping, Network Slicing, and Monitoring and Management are some of the uses of the Quality-of-service functions in optical network capacity maximization [68,69].

b. Cloud-based Solution

Cloud-based solutions have become a potent method for telemedicine networks to operate as efficiently as possible. Telemedicine providers may increase patient care, save costs, and increase access to healthcare services by utilizing the power of the cloud through the optical telecommunications infrastructure. These are several strategies for telemedicine networks that can employ cloud-based solutions to increase efficiency [70]. Cloud-based solution for telemedicine infrastructure is depicted in Figure 5.

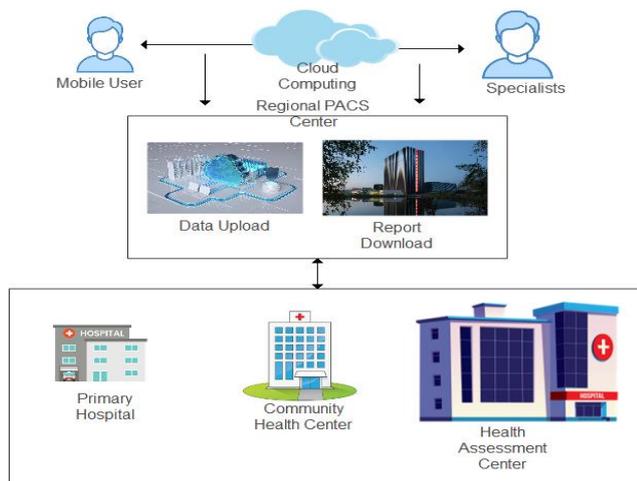

**Figure 5:** Cloud-based solution for telemedicine infrastructure

Scalable resources provided by cloud-based technologies allow telemedicine operators to swiftly and simply expand or decrease resources in response to demand. This guarantees that medical facilities can satisfy the rising demand for telemedicine services without sacrificing quality or performance. Moreover, cloud-based solutions let telemedicine service providers combine several telemedicine platforms, like messaging, video conferencing, and remote monitoring [71]. This increases efficiency and lowers costs by enabling healthcare practitioners to offer a variety of telemedicine services from a single platform.

c. Machine Learning (ML) approach

Machine learning and artificial intelligence (AI) algorithms can be used with cloud-based systems to evaluate vast volumes of patient data and offer in-the-moment insights into patient health. By doing so, healthcare professionals can better understand potential health hazards and treat patients. By reducing latency and optimizing network efficiency, ML and AI algorithms can raise the quality of telemedicine services. Predictive algorithms that adapt network settings in real-time based on the type of data being transmitted can help with this. Critical telemedicine applications like video conferencing and remote monitoring can be given the bandwidth and low latency they require to operate at their best using ML and AI to prioritize network traffic [72]. Moreover, it can be used to dynamically assign bandwidth based on network demand, ensuring that resources are used effectively and efficiently. This can aid telemedicine providers in cutting expenses and enhancing network efficiency.

Some other salient techniques that have been used to improve the optical network capacity for telemedicine are the use of Wavelength Division Multiplexing (WDM) [73], Optical Amplifiers [74], Dynamic Optical Networks (DON) achieved with the use of multiplexers and SDN [75]. These discussed considerations are important to consider whilst designing a telemedicine network. However, the physical layer, the network layer, as well as the operational and commercial requirements of the network must all be taken into account to maximize optical network capacity for telemedicine. This may entail a number of technologies and methods as well as continuing network maintenance and optimization.

## 5.0 DESIGNING A TELEMEDICINE NETWORK

The designing of a telemedicine network in this subsection focused on a design that can improve network capacity, control, and management. The Internet of Things (IoT) technology has been the subject of numerous studies with the aim of making telemedicine smart. These study fields include cognitive radio network communications, artificial intelligence methodologies, and techniques, Internet of Things wearable sensors and hardware devices, smartphones, and cloud computing, which all aim to improve telemedicine architecture and design. The following subsections will discuss different telemedicine network architecture design approaches.

## 5.1 Cognitive Radio-Based Telemedicine Network Design

Researchers' interest in the wireless transfer of patient medical data via diverse networks has lately increased. Yet, challenges like efficient frequency spectrum usage and longer gadget lifespans are thought to be the most urgent ones right now [76]. MBANs are now being used in unlicensed frequency bands, where there is a high risk of cross-interference with other electronic devices [77]. Feng, et al., colleagues (2010) investigated the effectiveness of enabling telemedicine traffic in a CRN to address the aforementioned issue. Procedures and performance for enabling urgent and real-time periodic telemonitoring were proven after the installation of an infrastructure-based CRN for telemedicine. Traffic on the network suggests that telemedicine traffic can obtain sufficient real-time performance [78]. [77] suggests a workable architecture for an MBAN based on UWB radio technology with useful CR capabilities. In this scenario, cognitive capacities are developed via multiband orthogonal frequency-division multiplexing and impulse radio (IR).

The key technological contribution of the work is the design of an MBAN with frequency agility and frequency domain spectrum shaping capabilities. When multiple devices are operating in the same frequency segments and are placed close to one another, as may occur in busy medical facilities, these qualities make it simpler to prevent interference. A Bio Cog algorithm was also developed to set up cognitive networks for disseminating medical data. Using a range of wireless technologies, such as XBee, Wi-Fi, and Bluetooth [76], it uses an efficient frequency spectrum allocation approach that takes network heterogeneity into account. [79] design a three-tiered telemedicine platform to offer prompt medical services. Body sensors serve as the first layer of the design, while a UWB communication-enabled gateway serves as the second tier. In the third tier, the hospital and the mobility vehicle were connected via cognitive radio-enabled technology.

Network connections are often irregular and occasionally unimpressive in a remote or rural place in terms of bandwidth availability. For the e-healthcare system to be able to adjust its transmission process based on the status of the communication network, an adaptive network architecture must be developed. In order to address network congestion and irregularity, as well as to provide priority-based health services to outlying primary health care facilities, a dependable wireless telemedicine network for e-health applications was proposed and assessed using a cognitive radio network technology. A MAC protocol that handles emergency data was created to order patient data in accordance with their medical conditions. By switching the data transmission process to any of the available networks (GSM, 3G-UMTS, WiMAX, and 4G-LTE), the framework was able to address network congestion and consistency difficulties [80].

[81] suggested a particle swarm optimization approach for essential medical wireless application networks employing cognitive radio-enabled WBAN for effective battery use and uninterrupted data transfer in BAN. For the conveyance of patient data inside hospitals and telemedicine, [82] suggests an overlay cognitive radio method. In this case, the primary user is a set of telemedicine transmitters and receivers, and the secondary user is an in-hospital patient monitor. The bit error rate and spectral efficiency can both be improved with this technique.

## 5.2 Artificial Intelligent Based Telemedicine Network Design

Two recent technology developments, artificial intelligence, and robots, may be helpful in telemedicine, in enhancing the human ability to respond to pandemics and other catastrophes, and in supporting home-based care. While developing decision support systems, more care must be given to applications in the medical industry [83]. In order to avoid cancer misdiagnosis caused by human error or improper data interpretation, [83] presents a reliable computer-aided diagnosis method supported by intelligence learning models. To improve prediction performance, a feature modeling approach based on machine learning is suggested. The investigations made use of datasets relevant to breast, cervical, and lung cancer. Using supervised learning methods, the best characteristics minimized by the proposed system are trained and validated [83].

[84] recommend a logical-centered software design for popular health monitoring apps. To swiftly connect third-party sensor devices in BAN, PAN, and NAN Area Networks, the system features a transparent, smartphone-based sensing architecture with adaptable wireless interfaces and plug-and-play capability. By combining different components and scripting the application logic, the new visual Inference Engine Editor enables machine learning

specialists and medical specialists to build data processing models.

[85] gives a prediction model for a mobile phone-based video streaming system that was developed and tested for dementia patients using a number of carefully selected data mining approaches for categorization. The aptitude, living conditions, and preferences of each individual were taken into consideration. It is now possible to collect physiological data from mobile patients using data-collection systems, but the majority of these tools generate so many false-positive alarms that they cannot be employed in routine clinical practice. Wearable patient monitors are used by [86] to assess vast quantities of continuously collected, multivariate physiological data utilizing guiding machine learning algorithms. The purpose is to provide early notice of significant physiological determinants so that some form of predictive therapy can be administered.

[87] proposed a system for ankle rehabilitation that combines several technologies, including 3D printing, machine learning, and a smartphone (the iPhone) wireless gyroscope platform. The majority of ankle rehabilitation system is made using 3D printing. An email attachment from a smartphone's wireless gyroscope platform records the therapy sessions used with an ankle rehabilitation device. Data from the gyroscope signal is prepared for machine learning. When using the ankle rehabilitation system, a support vector machine can accurately distinguish between an ankle that has been affected by hemiplegia and an unaffected ankle in 97 percent of cases. [88] provides a personal health platform that makes it possible for an Android device to gather, analyze, and communicate sensor data to observation storage using interoperability standards. The acquired data was compressed using compressed sensing techniques, and the genetic algorithm was used to optimize it.

[89] proposes a new paradigm for healthcare services (HCS) in a cloud environment and employs Parallel Particle Swarm Optimization (PPSO) to enhance VM selection. A brand-new model for identifying and predicting chronic kidney disease (CKD) is also released in order to evaluate the efficacy of the virtual machine model. The CKD prediction model was developed using two subsequent methodologies: linear regression (LR) and neural networks (NN). It is possible to pinpoint significant factors that have an impact on CKD using LR. NN is used to predict CKD. The outcomes show that the suggested method considerably improves system efficiency for real-time data retrieval and cuts down on overall execution time. Geotagged tweets from Twitter, statistics on

influenza-like illnesses from the Centres for Disease Control and Prevention, and an algorithm designed for artificial neural networks were also utilized to forecast disease in real time [90]. These geotagged tweets highlighted the location from which the user posted them, enabling the Twitter App to track their location. A network-based approach that takes advantage of spatiotemporal trends in previous influenza activity was used to predict the epidemiology of influenza in the USA each season. This approach combined a predictive method of self-correction with Google Patterns relevant to influenza, cloud-based EHRs, and historical flu trends. [91].

[92] investigates several robotic and artificial intelligence-based telehealth applications during COVID-19 and suggests a unique artificial intelligence-assisted telemedicine architecture to hasten the adoption of telemedicine and broaden access to high-quality, inexpensive healthcare. [93] proposed presents an efficient and automated method for diagnosing epileptic seizures that combines level-crossing sampling and adaptive-rate processing. Utilizing a level-crossing analog-to-digital converter (LCADC), the electroencephalogram (EEG) signal was obtained, and its active parts were chosen using an activity selection algorithm (ASA). The method's performance in classifying epilepsy also serves as a measure of the method's overall accuracy. For the majority of the analyzed examples, the suggested approach achieves 100% classification accuracy. In the United States, deep learning recurrent neural networks (RNNs) were employed to predict influenza outbreaks at province and municipal scales, with the help of bioinformatics methods like docking and modulation to forecast forthcoming influenza subtypes that could trigger a future pandemic [94].

### 5.3   IoT-Based Telemedicine Network Design
Many studies on the Internet of Things (IoT) technologies have been conducted with the goal of enhancing telemedicine networks and intelligence. The IoT-based telemedicine network is one of the crucial elements for helping society prevent illness and provide high-quality healthcare services [95]. In order to support e-health applications and address network security and packet loss issues in IoT-based body area networks, [96] developed a 5G network design. This is done by caching network content using a content-centric network router and exploiting the efficient resource management, traffic reduction, and scalability of the 5G network. For IoT-based smart medical systems, [97] presents a distributed coexistence mitigation technique that consists of two steps (the channel planning stage and the medium access adjustment stage), both of which can dynamically minimize interference in the coexistence state.

[98] propose an Internet of Things (IoT) LoRa-based tracking and monitoring system for patients with a mental illness. The system is made up of LoRa gateways that are situated in public areas like hospitals as well as a LoRa client, which is a tracking device that is installed on the patient's end device. The LoRa gateways are linked to local and cloud servers using mobile cellular and Wi-Fi networks as the communication medium. [99] explains a Fog architecture that used big data analysis and unsupervised machine learning to discover patterns in physiological data. A prototype was developed and evaluated using aberrant speech data from people with Parkinson's disease who were being watched in real life (PD). The suggested architecture analyzed data on aberrant speech obtained from smart watches worn by Parkinson's disease sufferers using machine learning. In [100], Four typical IoT components were designed as part of a wearable 12-lead ECG Smart Vest system for the early diagnosis of cardiovascular illnesses (CVDs). The first component is a sensor layer made of a textile dry ECG electrode, and the second is a network layer using Bluetooth, Wi-Fi, and other technologies. The next elements consist of a server calculation and a cloud-saving platform. The application layer, which comes last, is in charge of signal analysis and decision-making.

A low-cost modular monitoring system prototype was developed in [101] using arrays of low-power EKG, SpO2, temperature, and movement sensors. Its objective is to offer mobile support so that in an emergency, medical care can be delivered more quickly and effectively. The IoT concept was used to construct the interfaces for these sensors: In order to ensure platform independence and enable developers a flexible approach to add new components, a central control unit provides a RESTful-based Web interface. from the authors of [102], an online telemedicine system is built using relay selection algorithms. Depending on the relay node selection method, one or more network nodes may be utilized. According to the effectiveness of the provided method, the system throughput will increase, nodes won't be used repeatedly, and the system will be more stable. [103] demonstrated a smart healthcare platform based on cutting-edge technologies including machine learning and the Internet of Things (IoT). The technology is intelligent enough to perceive and understand a patient's data using a medical decision support system, allowing people in remote locations to utilize it to assess whether they have a serious health issue and treat it accordingly by phoning neighboring hospitals.

The authors in the work of [104] addressed the fundamental issues of healthcare service provision when frequent failures in a telemedicine architecture occur, a newly distributed fault-tolerant mHealth framework-based Internet of things (IoT) is developed in this study. To support the development of chronic heart disease (CHD) via telehealth in a remote setting, two models are provided. Hocine Hamil et al, [105] created a new telehealth system that uses the e-Health sensors platform, XBee modules, and Arduino Uno and Raspberry Pi as the acquisition and processing units, respectively, and allows for the categorization of numerous bio signals and protected wireless transmission. To monitor the patient's health status, threshold detection can be used to assess the acquired data on things like temperature, Galvanic skin reaction, and blood oxygen levels. By classifying Electrocardiogram (ECG) data using Artificial Intelligence (AI) by utilizing TensorFlow and Keras tools, the prediction of the cardiac condition based on automatic recognition of arrhythmias is proven by the authors.

From [106], a smart monitoring and emergency alarm system for COVID-19 patients was suggested as an Internet of Things (IoT) design. A temperature sensor, a blood oxygen level sensor, and a heart rate sensor would be used in this system to monitor a patient who is in stage 1 of the condition. An Arduino Uno Controller would also be used to gather data from the patient and transmit it to an IoT server. The authors believe that the suggested technique will be dependable and successful in reducing mortality and hospital admissions by forewarning and saving lives in times of crisis. When the patient is in a region with extreme conditions, the device is also intended to transmit message warnings to the nearby hospital.

### 5.4 Mobile Phone base Telemedicine Network Design

Mobile telemedicine is a research area that makes use of advancements in cellular telecommunications networks to give extremely flexible medical services that are not possible with the conventional telephone. [107] describes the design of a prototype integrated mobile telemedicine system that is interoperable with current mobile telecommunications networks as well as third-generation networks. A physician ought to be permitted to remotely monitor a patient who is free to move around for sports medicine and emergencies thanks to technology. [108] designed a prototype emergency telemedicine system that uses a CDMA 1X-EVDO reverse connection to transmit both biological signals and patient motion video from a moving vehicle. The application layer protocol of the designed prototype system included frame rate control using MPEG-4 compression, error control using automatic repeat request, and priority control between the vital sign and video images to address the limited bandwidth of the reverse link (transmission bandwidth of cellular devices) in comparison to the forward link (receiving bandwidth). In [109] Create a cutting-edge telemedicine system for urgent care. A team in a small rural hospital can be remotely led by an emergency care specialist from a large referral hospital when treating critically ill patients thanks to the system. It features a seamless interface to the intricate clinical working environment and allows for the transmission of high-quality audio/video information.

For remote locations, [110] suggested telemedicine and a Clinic-In-A-Can (CIAN). CIAN is a portable medical institution that may be set up as an emergency healthcare infrastructure anywhere in the world. It's built within a shipping container. The model includes a web-based telemedicine system that provides the essential features required for audio and video conferencing-based medical teleconsultations. Communication between nurses and doctors, nurses and community health workers (CHWs), and doctors and patients will be made easier by technology. [111] employs a system that combines long-term care, constant monitoring of a number of critical indicators, and an emergency cellular link to a hospital. Normally, the internet is used to transfer all of the raw data that has been gathered. The proposed system may continually gather four physiological signals, such as blood pressure, temperature, SpO2, and ECG, and transmit those signals to an intelligent data processing system to identify abnormal pulses and look into possible chronic conditions. The work of [112] established a real-time emergency telemedicine system for remote medical diagnosis and demonstrate the feasibility of performing haematological tests in an ambulance using accurate real-time wireless transmission to the referral hospital. [113] created an intelligent sensor-based telemedicine system that can identify and assist diabetes patients. Using a back-propagation technique, training data, and a neural network feed-forward prediction model, the system evaluates whether the patient is at risk of acquiring diabetes.

Using a range of information technologies, [114] designed and constructed a web-based telemedicine system that enables patients to get medical consultations remotely. It also offers long-term medical staff monitoring of the patient's physical condition to satisfy non-contact requirements, i.e., the system platform exchanges opinions between the two parties, in addition to providing feedback regularly to the system platform for self-measurement of physiological conditions.

Based on the different aspects of telemedicine network design and with respect to other support for telemedicine networks with a view to maximizing network capacity, some requirements identification as well as recent challenges will be presented subsequently. However, technical challenges in telemedicine network design vary with respect to the deployment platform and applications.

## 6.0 RECENT RESEARCH CHALLENGES
This subsection presents the identified challenges in general with possible solution in obtaining efficient network capacity, control, and management in designing a telemedicine network. This is as shown in Table 4.

**Table 4:** Summary of various challenges for using UAV, SDN and ML to maximize network capacity, control, and management in designing telemedicine network

| No | Name of the challenge | Reason for consideration | Probable solution |
|---|---|---|---|
| 1 | Requirements Identification | Choice of solution depends on understanding the needs of all network participants in detail. | Identify the stakeholders<br>Conduct needs Assessment<br>Determine the technical requirements<br>Define network architecture as suggested by [116] |
| 2 | UAV Limited Payload Capacity | Unless acting only as a relay node, most UAVs have limited payload capacity, which can be a hindrance when transporting medical equipment or supplies | Employ several drones for various tasks, such as one for equipment transportation and another for telemedicine consultations and another as relay nodes. |
| 3 | UAV Data Transmission and Capacity | UAVs might face difficulties in area coverage due to speed and battery life limitations or remote arears that have unreliable connectivity. | As presented in Table 3, To secure access in remote places, use cutting-edge communication technologies like satellite-based internet or mobile network expansion. Utilize data compression methods to reduce bandwidth consumption |
| 4 | Environmental Challenges | windy conditions, signal degradation, and obstructions in the UAV's flight path | Use alternative stabilizing algorithms, more wing-equipped UAVs. |

| 5 | Security and privacy with the use of UAV-SDN and ML as a support to telemedicine | Because SDN introduces centralized control, the network may be more susceptible to hacker attacks and data breaches. Machine learning models might need access to private medical information, posing issues with data security and privacy. | Implement strong security mechanisms, such as access restriction, authentication, and encryption. Update and patch network controllers and devices on a regular basis. Ensure adherence to laws governing the protection of healthcare data, such as HIPAA. |
|---|---|---|---|
| 6 | Network Scalability | As the number of connected devices and remote consultations rises, telemedicine networks must be able to handle growing data loads. | SDN infrastructure should be scalable while being designed. Utilize SDN tools that are cloud-based and can dynamically assign resources in response to demand. Quality of Service (QoS) policies should be implemented to give priority to vital traffic. |
| 7 | Interoperability | Medical equipment and telemedicine applications may employ many communication protocols. | Use SDN controllers and gateways that can bridge between different protocols and support a variety of them. Encourage the adoption of industry standards for telemedicine equipment interoperability |
| 8 | Real-Time Performance Requirements | To provide real-time interactions between medical personnel and patients using telemedicine applications, low latency and high-quality connections are required. | Implementing QoS standards that allot enough bandwidth and low-latency channels for telemedicine traffic will prioritize real-time traffic. Optimize routing choices for low latency by using adaptive algorithms in machine learning models. |
| 9 | Algorithm Bias | Biases that are inherited by machine learning algorithms from training data may result in unfair treatment or decisions in telemedicine network. | Ensure that training data is diverse and representative to minimize bias. |
| 10 | Data Volume and Processing | Large amounts of data are produced by telemedicine, and machine learning demands a lot of computer power. | As presented in the diagram in Figure 5, To manage massive datasets and model training, use cloud-based solutions and distributed computing. Reduce latency and handle data more quickly by using edge computing. Employ edge computing to reduce latency and process data closer to the source. |

## 7.0 CONCLUSION

The use of Telemedicine applications and infrastructure is getting popular in today's world and becoming ubiquitous in many urban, remote, and disaster and rescue mission activities. Telemedicine or e-health or e-medicine is an applications field where medical practitioners have remote access to their patients to deliver adequate medical care. The telemedicine system not only allows communication of patients with doctors, but it also plays a very significant role in the diagnosis, management, and follow-up of patients across continents. This is not to say that telemedicine has not been practiced during the non-era of the internet but with the coming of the internet and high-speed networks, ideas, and existing technology-based applications are making major changes in medical science, which are in the aspect of e-health, rural telehealth care programs, home health care services, continuing medical education, outdoor patient education, international healthcare services, mobile medical services and post-disaster medical management amongst others.

Nonetheless, these services are only possible with a good network infrastructure. Therefore, the adoption of the optical network as an infrastructure to telemedicine network backbone by the research community can be attributed to the connectivity speed it possesses and the long-distance communication it offers amongst others. More so, health care providers and their patients have significantly more flexibility and communication options as a result of the fast-optical network, which ensures that medical institutions may transfer healthcare data securely and efficiently.

Detailed literature reviews have revealed significant study gaps in this area that need to be filled in thoroughly in order to make a greater contribution. As a result, the following list of potential future research areas is provided:

- To carry out studies on optimizing the telemedicine network to accommodate more remote users using UAVs as support.
- To study the benefits and challenges of AL and ML, applied to telemedicine networks using UAVs, to ease data processing and management, patient monitoring, medical extension, etc.
- To research to improve the network capacity and run-time of the UAV used to support telemedicine by optimizing the flight schedule, battery size, etc. as it will be necessary to operate them while providing medical aid. Here, using renewable energy sources could be a great way to address this problem.
- Although optical networks are typically thought of as safe, telemedicine applications could potentially pose security problems. To assess the security threats connected to telemedicine applications and create mitigation techniques, more research is required.
- Though optical networks are a good way to carry data, it is unclear how economical they are compared to other options. To determine areas where cost savings can be made and to assess the cost-effectiveness of optical networks in telemedicine applications, more cost economic analysis can be conducted.

- To maintain QoS during peak usage times or when network congestion occurs. More research is needed to understand the QoS requirements of telemedicine applications and how optical networks can meet these requirements. One recent technique suggested by works of literature is the use of SDN.